\documentstyle[aps,epsf]{revtex}
\def\be{\begin{equation}}
\def\ee{\end{equation}}
\def\beq{\begin{eqnarray}}
\def\eeq{\end{eqnarray}}
\def\lsim{\:\raisebox{-0.5ex}{$\stackrel{\textstyle<}{\sim}$}\:}
\begin{document}

\draft

\title{Strong Phase Separation in a Model of Sedimenting Lattices} 
\author{Rangan Lahiri$^1$\cite{nomore}, Mustansir Barma$^1$\cite{a}, 
and Sriram Ramaswamy$^2$\cite{b}}

\address{$^1$ Department of Theoretical Physics, 
Tata Institute of Fundamental Research, 
Homi Bhabha Road, Mumbai 400 005, India; 
$^2$Centre for Condensed Matter Theory, Department of Physics, Indian
  Institute of Science, Bangalore 560 012, India\\}
%27 August 1998
\date{\today \quad TO APPEAR IN PHYS REV E, 1 JANUARY 2000}
\maketitle
\begin{abstract}

We study the steady state resulting from instabilities in crystals
driven through a dissipative medium, for instance, a colloidal crystal
which is steadily sedimenting through a viscous fluid. The problem
involves two coupled fields, the density and the tilt; the latter
describes the orientation of the mass tensor with respect to the driving
field. We map the problem to a 1-d lattice model with two coupled species
of spins evolving through conserved dynamics. In the steady state of this
model each of the two species shows macroscopic phase separation. This phase
separation is robust and survives at all temperatures or noise levels---
hence the term Strong Phase Separation. This sort of phase separation can
be understood in terms of barriers to remixing which grow with system size
and result in a logarithmically slow approach to the steady state.
In a particular symmetric limit, it is shown that the condition of detailed
balance holds with a Hamiltonian which has infinite-ranged interactions,
even though the initial model has only local dynamics. The long-ranged
character of the interactions is responsible for phase separation,
and for the fact that it persists at all temperatures. Possible experimental
tests of the phenomenon are discussed.

\end{abstract}

\section{Introduction} 
\label{intro}
\subsection{Background}
\label{background}
Sedimentation -- the settling of heavier
particles in a lighter fluid -- is a rich source
of intriguing physics \cite{blanc}.  The steadily
sedimenting state arises, of course, from a
balance between gravity and viscosity.  
Viscous damping in this nonequilibrium steady state 
has important consequences: when a given
particle is slowed down by the fluid, its  
momentum does not disappear, but produces
disturbances in the fluid which affect the motion
of other particles \cite{russel,happel}.    This
makes sedimentation a challenging problem in
the statistical physics of driven many-body
systems.  

In the general area of nonequilibrium steady
states, much recent progress has come by stepping
away from the difficulties of hydrodynamics and
focussing instead on simple driven
lattice-gas models \cite{ddlg}. 
In fact, intimate connections were
discovered by two of the present authors
(hereafter LR) \cite{rlsr} between these 
models and the physics of sedimenting crystalline
suspensions (as well as a closely related problem, 
a flux-point lattice moving through a
superconducting slab). The LR model was based on two
crucial properties of collective settling discovered 
by Crowley \cite{crowley} in his theoretical and 
experimental studies of hard spheres sedimenting 
in a viscous medium:  
(i) The {\em magnitude} of
the local settling velocity of a region of the
crystal depends on its {\em concentration}, i.e.,
on the particle number density in that region,
and (ii) the {\em direction} of the local
settling velocity depends on its {\em tilt}, that
is, the orientation, relative to the applied
force (gravity) of the principal axes of the
local particle distribution. These effects, 
which also follow from symmetry arguments, were
incorporated into a natural one-dimensional
model for the coupled, stochastic, local spin-exchange
dynamics of two sets of Ising variables,
$\{\sigma_i\}$ with states denoted by $+$ and $-$ 
for the concentration relative to the mean, and
$\{\tau_i\}$ with states denoted by / and
$\backslash$ for the tilt, on the sites $i$ of
a one-dimensional lattice. Analysis of this 
model leads to several
interesting results, some published in
\cite{rlsr} and some new ones which we present
here. 

\subsection{Update rules}
\label{updaterules}
Our results will be easier to understand after a quick summary of the
update rules of the lattice model, which we turn to next. This will
also serve to underline the simple nature and potentially wide
applicability of the model.  It is convenient to place the
$\{\sigma_i\}$ and $\{\tau_i\}$ on two sublattices of our one-dimensional
lattice; we label sites on the first sublattice by integers, and those
on the other by half-integers.  A configuration is then a string
$\tau_{1\over 2}\sigma_1 \tau_{3\over2} \sigma_2 \tau_{5\over2}
\sigma_3
\tau_{7\over2}\sigma_4 \ldots$, say $/ + \backslash - / - / +
\backslash \ldots$. Using the above notation for the
states of the two variables, and denoting the
rate of an exchange process by $W$, the probabilities per 
unit time for the various possible exchanges can be represented
succinctly by 
\begin{eqnarray}
\label{rates}
W(+ \backslash - \rightarrow - \backslash +) & = & D  +  a  \nonumber \\
W(- \backslash + \rightarrow + \backslash -) & = & D  -  a  \nonumber \\
W(- / + \rightarrow + / -) & = & D'  +  a'  \nonumber \\
W(+ / - \rightarrow - / +) & = & D'  -  a'  \nonumber \\
W(/ + \backslash \rightarrow \backslash + /) & = & E  +  b \nonumber  \\
W(\backslash + / \rightarrow / + \backslash) & = & E  -  b \nonumber  \\
W(\backslash - / \rightarrow / - \backslash) & = & E'  +  b' \nonumber  \\
W(/ - \backslash \rightarrow \backslash - /) & = & E'  -  b', \nonumber  \\
\end{eqnarray}
where the first line, for example, represents the rate of $+-$ going to $-+$ in 
the presence of a downtilt $\backslash$, and so on. The quantities $D, \, E, 
\, D', \, E'$ (all positive) and $a, \, b, \, a', \, b'$ are all in 
principle independent parameters but we will argue below that the case
of physical interest and relevance to the sedimentation and driven
flux-lattice problems is sgn$a =$ sgn$a'$, sgn$b =$ sgn$b'$, and that
the quantity which controls the qualitative behaviour of the model is
then $\alpha = {\rm sgn}(ab)$. We find two completely distinct kinds
of behaviour, depending on whether $\alpha$ is positive or negative.
If $\alpha < 0$, the steady state of the model is a mixture of pluses
and minuses, and of uptilts and downtilts, which is statistically
homogeneous on a coarse-grained level. If $\alpha > 0$, such a state
is {\em unstable} with respect to fluctuations which drive it to a
strongly phase separated state of a type defined and discussed
below. We refer to the cases $\alpha < 0$ and $\alpha > 0$ as the
Stable and Unstable LR (SLR and ULR) Models respectively.

\subsection{Strong Phase Separation: Summary of Results}
\label{spssum}
The focus of this paper is the study of phase 
separation phenomena of a new and unusual sort, in the unstable LR (ULR) 
model of sedimenting colloidal crystals described above. 
Following the appearance of the LR model,
the same type of phase separation was shown to occur 
\cite{c3cm} in a three-species permutation-symmetric model
on a one-dimensional lattice with periodic boundary conditions, with
dynamics which may be regarded as a simplified version of that in
\cite{rlsr}.  A further generalisation which breaks the permutation
symmetry between the three species was studied in \cite{rittenberg}.
The underlying mechanism of phase separation appears robust and simple
enough that it might be worth looking for in other systems. Here is a
summary of our results.
 
\begin{enumerate}
\item In the present context, phase separation involves   
the spontaneous formation of
macroscopic domains of $+$ and $-$ as well as / and
$\backslash$ in the ULR Model \cite{rlsr} . This segregation  
is robust in that it survives at all temperatures $T$. 
Let us recall that most statistical systems which
show phase separation at low $T$ (or low
noise-level, in nonequilibrium cases 
\cite{grinben,toom}) lose this property at higher
$T$ or noise strengths.  Certainly if
one were to think in terms of energy and entropy,
this would be the general expectation.  Against
this backdrop, a phase separation so robust as to
persist at {\em all} finite $T$, and in a
one-dimensional system at that, is quite
unexpected. We suggest the name Strong
Phase Separation (SPS) for this unusual 
phenomenon. 

The importance of SPS in the ULR model arises 
from the close relation of the latter 
to a physically realisable system of considerable current interest, 
namely, sedimenting colloidal crystals. Towards the end of 
this paper we suggest experiments which can be performed 
on fluidised beds of colloidal crystals to test some 
of the ideas presented in this work.  

\item The occurrence of SPS can be seen best in a certain 
limit in which the dynamics of the ULR model obeys the  
condition of detailed balance. In this limit, an energy function $E$ can
be constructed such that the steady-state
probability of a configuration $\{\sigma_i,
\tau_{i-{1\over 2}}\}$ is proportional to $\exp\left[- E(\{\sigma_i,
\tau_{i-{1\over 2}}\}/T)\right ]$. 
Although the dynamics is entirely
local and involves rates of order unity, the emergent
energy function $E$ for the effective equilibrium
theory involves interactions of unbounded range. 
As a result, $E$ has a nonextensive (more properly, superextensive)
character, which is how our model and those 
of \cite{c3cm,rittenberg} manage  
to get around the usual obstacles \cite{ll} to
phase separation in one dimension. In our model, $E$ has a simple 
interpretation: it is the energy of a collection of particles, {\em viz}. the 
$\sigma_i$s, in a potential landscape built from $\{\tau_{i-{1\over 2}}\}$. 
The superextensivity is then a consequence of having potential 
energy wells whose depths scale with the system size.   

Thermodynamic properties can be calculated 
in the strongly phase-separated state. In particular,   
the width of the interfacial region is found to vanish as 
$T \rightarrow 0$ and diverge as $T \rightarrow \infty$.  

\item Strong phase separation is a robust phenomenon, and 
persists even when the condition of detailed balance 
does not hold. This can be seen through arguments \cite{rlsr}  
based purely on kinetics without recourse to 
an energy function: the transport of a $+$ from 
one end of a $+++...+$ domain to a point a distance 
$n$ away requires a time which grows exponentially 
with $n$, as $n$ moves against the tilt field would 
be required. Thus a macroscopically phase-separated 
state would be expected to survive infinitely long 
in the infinite size limit.  

\item Although phase separation is inevitable in
the ULR Model \cite{rlsr} and in the models of \cite{c3cm,rittenberg}, the kinetics of domain
growth is anomalously slow. The barriers that
oppose the remixing of the macroscopically
segregated state also inhibit the processes of
diffusion that cause large domains to grow at the
expense of smaller ones. These barriers,
moreover, are produced by the dynamics of the
model, not introduced {\em ex machina} in the
form of quenched randomness. This results
in intriguing {\em ageing} effects: for instance,  
the growth of domains is logarithmic in time, 
as has been verified in numerical studies in 
\cite{c3cm}. Further, in the detailed-balance limit,
the decrease of $E(\{\sigma_i, \tau_{i-{1\over 2}}\})$ is logarithmic
in time as well. Thus, despite the {\em existence} of a thermodynamic
equilibrium state in the detailed-balance limit, a system which starts
from a random initial condition has an extraordinarily difficult time
reaching it. Such a system is best thought of as perpetually evolving, never in a
truly steady state, sinking slowly into progressively deeper minima,
in a manner which recalls the glassy state of the model of
\cite{bouchaud}.

\item Arguments given in \cite{rlsr} already amounted to
showing that SPS occurred in the ULR model. 
Specifically, it was shown there that 
the remixing of phase-separated domains in
would {\em always} be
opposed by barriers whose height diverged with
the system size. The simulation results of \cite{rlsr}, 
however, were complicated by the presence of 
a repulsion between adjacent $+$ sites, 
which modelled interactions between charged 
colloidal particles. Increasing this repulsion beyond 
a threshold value led, in the numerical studies of \cite{rlsr}, 
to an apparent loss of phase separation. It is now clear, from 
the calculations reported in the present paper, that the
observed remixing \cite{rlsr} was a finite-size effect. 
\end{enumerate}

\subsection{Outline}
The rest of this paper is organised as follows.
In the next section (\ref{contmodel}), we review the
derivation \cite{rlsr} of continuum equations of
motion for a crystalline array moving through a
dissipative medium, and show how, at the
linearised level, they lead to either a new class
of ``kinematic waves'' \cite{lighthill} or an
instability towards phase separation. Section \ref{contmodel}  
closes by presenting a simplified one-dimensional
continuum model which retains all the essential
features of the higher-dimensional problem. In
section III, we use arguments similar to those
connecting the noisy Burgers equation to the
driven diffusive lattice gas \cite{ddlg} to
construct the LR lattice-gas model \cite{rlsr}
whose long-wavelength limit has the relevant physics 
of the aforementioned one-dimensional continuum
equations. We show, in a certain highly 
symmetric limit, that the Unstable LR model has  
a detailed balance property.  In this limit we demonstrate
Strong Phase Separation and calculate thermodynamic 
quantities. Further, we give arguments to show that 
SPS occurs in the entire parameter range of 
the ULR model.  We argue that the coarsening of
domains in the ULR model is ultraslow, with a
characteristic length scale growing
logarithmically in time.  An
analysis of a continuum model for SPS  
is the subject of section IV. Section V summarises 
our results and suggests experiments to test our 
predictions. 

\section{Continuum dynamical model for a moving crystal} 
\label{contmodel}
\subsection{Motivation}
\label{contmodelmotiv}
The LR lattice-gas model \cite{rlsr} arose as a
simplified description of the dynamics of a
crystal moving steadily through a dissipative
medium.  It is therefore useful to review the
construction of the continuum equations of motion
for such a system. There are at least two
physical situations where this dynamical problem
arises: (i) the steadily sedimenting colloidal
crystal mentioned above; (ii) a flux-point
lattice moving through a thin slab of type
II superconductor under the action of the Lorentz
force due to an applied current. In (ii), the
dissipation comes both from the normal core of
the vortices and from disturbances in the
order-parameter and electromagnetic fields in the
region around the vortices. There is in principle
an important difference between the sedimentation
and moving flux-lattice problems: in the former,
the disturbances produced by the moving crystal
are carried to arbitrarily large length scales by
the long-ranged hydrodynamic interaction, while
in the latter, both electromagnetic and
order-parameter disturbances are screened and
are thus limited to a finite range. A complete analysis of
the sedimentation dynamics of a three-dimensional
colloidal crystal thus requires the inclusion of
the hydrodynamic velocity field as a dynamical
variable. Instead, we  
consider an experimental geometry in which a
thin slab of colloidal crystal (with 
interparticle spacing $\ell \gg$ particle size) is confined
to a container with dimensions $L_x, L_z \gg L_y
\sim \ell$ (gravity is along $-\hat{\bf z}$). The
{\em local} hydrodynamics that leads to the
configuration-dependent mobilities
\cite{crowley,rlsr} is left
unaffected by this, but the long-ranged
hydrodynamic interaction is screened in the $xz$
plane on scales $\gg L_y$ by the
no-slip boundary condition at the walls. The model equations
(\ref{eom}) in dimension $d=2$ apply to such a system. 

\subsection{Constructing the equations}
Our construction of the
equations of motion ignores inertial terms, which is
justified both for the confined colloidal crystal
and, except at very low temperatures \cite{dmgtvr}, for the
flux lattice. Rather than keeping track of
individual particles, we work on scales $\gg \,
\ell$, treating the colloidal crystal or flux
lattice as a permeable
elastic continuum whose distortions at point $\bf
r$ and time $t$ are described by the (Eulerian)
displacement field ${\bf u}({\bf r},t)$. In
general, the equation of motion in the completely
overdamped limit has the form velocity =
mobility $\times$ force, i.e., 
\begin{equation}
{\partial \over \partial t} {\bf u} =
{\mbox{\boldmath $\mu$}}(\nabla {\bf u}) ({\bf
K}\nabla \nabla {\bf u} + {\bf F} + {\bf f}). 
\label{eomgeneral}
\end{equation}
In (\ref{eomgeneral}), the first term in
parentheses on the right-hand side represents
elastic forces, governed by the elastic tensor
$\bf K$, the second ($\bf F$) is the applied
force (gravity for the colloidal crystal and the
Lorentz force for the flux lattice), and $\bf f$
is a noise source of thermal and/or hydrodynamic
origin. Note that in the absence of the driving
force $\bf F$ the linearised dynamics of the displacement
field in this overdamped system is purely {\em
diffusive}: $\partial_t u \sim \nabla^2 u$, with
the scale of the diffusivities set by the product
of a mobility and an elastic constant. All the
important and novel physics in these equations,
when the driving force is nonzero, lies in
${\mbox{\boldmath $\mu$}}$, the local mobility
tensor, which we have allowed to depend on
gradients of the local displacement field. The
reason for this is as follows: The damping in the
physical situations we have mentioned above
arises from the interaction of the moving particles with the
medium. A dynamical friction of this kind will in
general depend on the local arrangement of
particles \cite{crowley,aditi}. Even for a perfect, undistorted
lattice, the symmetry of the mobility tensor will
thus reflect the symmetry of the underlying
lattice. If the structure in a given region is
distorted relative to the perfect lattice, the
local mobility will depart from its ideal
structure as well. Deviations of the structure
from the perfect crystal are described by the
full distortion tensor $\nabla {\bf u}$
\cite{elast} rather than its symmetric part, the strain,
since we are not in a rotation-invariant
situation. We further make the reasonable
assumption that the mobility can be expanded in a
power series in the distortion:
\begin{equation}
{\mbox {\boldmath $\mu$}} ({\bf \nabla u}) =
{\mbox {\boldmath $\mu_0$}} 
+ {\bf A } {\nabla \bf u} + 
{\cal O} (({\nabla {\bf u})}^2), 
\label{mobility}
\end{equation}
where {\boldmath $\mu_0$} is the 
mean macroscopic mobility of the undistorted
crystal. 

For a $d$-dimensional crystal driven steadily along the $z\/$
direction, assuming isotropy in the
$d-1$-dimensional ``$\perp$'' subspace normal to
$\hat{\bf z}$,  but {\em not}
under $z \rightarrow -z\/$, (\ref{eomgeneral})
and (\ref{mobility}) lead directly to 
\begin{mathletters}
\begin{eqnarray}
\dot {{\bf u}_{\bot}}& = &{\lambda}_1 {\partial}_z {\bf u}_{\bot} +
{\lambda}_2 {\nabla}_{\bot}  u_z
\nonumber
\\
&&+ {\cal O} ( \nabla \nabla u)
+ {\cal O} ( \nabla u  \nabla u) + {\bf f}_{\bot},
\label{eom1}
\\
\dot { u_{z}}& = &{\lambda}_3 {\nabla}_{\bot}.{\bf u}_{\bot} +
{\lambda}_4 {\partial}_{z}  u_z
\nonumber
\\
&&+ {\cal O} ( \nabla \nabla u)
+ {\cal O} ( \nabla u  \nabla u) +  f_z, 
\label{eom2}
\end{eqnarray}
\label{eom}
\end{mathletters}

\noindent where the constant drift along $z$ has been removed by 
shifting to the mean rest frame of the crystal. The terms that are
manifestly most important at small wavenumbers, at least within a
linear description, are the linear, first-order space derivative
terms. These terms arise from (\ref{eomgeneral}) and (\ref{mobility})
via the leading distortion-dependence of the mobility tensor,
multiplied by the driving force $F$. The coefficients ${\lambda}_i$
[as well as those of the ${\cal O} (\nabla u \nabla u)$ terms, as can
be seen from (\ref{eomgeneral}) and (\ref{mobility})] are thus
proportional to $F$, and the corresponding terms are therefore present
only in the driven state.  At small enough wavenumbers ($\lsim F/K$
where $F$ is the magnitude of the driving force density and $K$ a
typical elastic constant), these terms dominate the diffusive terms
coming from the elasticity. The terms of this type in (\ref{eom1})
tell us that a tilt (a $z$ derivative of a $\perp$ displacement or a
$\perp$ derivative of a $z$ displacement) leads to a lateral drift,
and those in (\ref{eom2}) imply that the vertical settling speed
depends on the compression (or dilation).  Since the system is not
invariant under rotations, there are no grounds for insisting that
${\lambda}_1 = {\lambda}_2$ or ${\lambda}_3 = {\lambda}_4$. ${\bf f}$
is a spatiotemporally white noise source containing the effects of
thermal fluctuations as well as chaotic motion due to the hydrodynamic
interaction
\cite{chaos,sedliq}. The reader will note that the form of the
diffusive second derivative terms and the distortion-dependence of the
mobility beyond linear order has been left rather general. This is
because even for $d = 2$, as can be seen by exhaustive listing,
symmetry under $x
\rightarrow -x, \, u_x \rightarrow -u_x$ permits,
all told, in (\ref{eom1}) and (\ref{eom2}), ten 
terms (this counting was wrong in \cite{rlsr}) bilinear in $\nabla {\bf u}\/$ and six
linear second derivative terms, with as many
independent coefficients. It is clearly  
difficult to make very useful general statements
about a problem with so many phenomenological
parameters so we restrict ourselves, in the next
subsection, to a linearised description to
lowest order in gradients.  We will return to the effects of
nonlinearities in later subsections. 

\subsection{Mode structure} 
If we retain only terms linear in the fields and
work only to leading order in wavenumber, then the relation
between frequency $\omega$ and wavevector ${\bf
k}$ implied by (\ref{eom}) is
\begin{equation}
\omega = {-1 \over 2} \left[(\lambda_1 +
\lambda_4) k_z \pm \sqrt{(\lambda_1 -
\lambda_4)^2 k_z^2 + 4 \lambda_2 \lambda_3
k_\perp^2} \right]. 
\label{omegakd}
\end{equation}
The dispersion relation (\ref{omegakd}) has a
wavelike character in all directions if
$\lambda_2 \lambda_3 > 0$. For $\lambda_2
\lambda_3 < 0$, while it is still wavelike for
$k_\perp = 0$, it has a growing mode
$\omega \propto -i k$ for $k_z \ll k_\perp$. 

{\em Linearly stable case --- kinematic waves}:
The wavelike modes are the generalisation, to the
case of a moving lattice, of the kinematic
waves which Lighthill and Whitham
\cite{lighthill} discussed in the context of
traffic flow and flood movements.  
The important difference in the
present case is that the waves propagate not only
along but also transverse to the direction of
drift. Some remarks towards a more complete
consideration of their dispersion relation,
including the effects of nonlinearities, may be
found in the context of a one-dimensional reduced 
model in \cite{rlsr}.  

{\em Linearly unstable case --- clumping}: In the
case $\lambda_2 \lambda_3 < 0$, for wavevectors
pointing outside a cone around the $z$ axis, the
system is linearly unstable, as already noted in \cite{rlsr}: 
small perturbations
grow, with a growth-rate which is {\em linear} in
their wavenumber. Whereas the linearised 
treatment cannot give detailed information about 
the final state of the system, we expect the growing mode 
to appear as a clumping and tilting of 
the colloidal crystal, with material concentrated 
at the bottoms of the tilted regions. The 
wavevector of the inhomogeneity will be 
mainly normal to the sedimenting direction.  

The remainder of this paper is directed towards 
a more detailed understanding of the statistical 
mechanics and dynamics of macroscopic clumping.  
Our studies are based mainly on the simplified one-dimensional 
lattice model of \cite{rlsr}. The construction of the 
lattice model is reviewed  
in section \ref{LRmodel}: 
its origins lie in a reduced, one-dimensional 
version of equations (\ref{eom}) which we now present. 

\subsection{A one-dimensional effective model}
\label{cont1d}
We saw above that the equations of motion for a
moving lattice contained terms of a qualitatively
new form, not present in the equations of a
lattice at equilibrium. To linear order, these were the
$\{\lambda_i\}$ terms in (\ref{eom}), which are  
proportional to the driving force, and of lower
order in gradients than those arising from the
elasticity of the crystal. The effects of the
linear instability for $\lambda_2 \lambda_3 < 0$ thus cannot be mitigated by
including the diffusive terms arising from the
linear elasticity. To see what final state, if
any, emerges from the initial unstable growth in
the case $\lambda_2 \lambda_3 < 0$, we must go
beyond a linear treatment. Even in the stable
case $\lambda_2 \lambda_3 > 0$, the combined effects of nonlinearities and
noise could result in effective dispersion
relations for long wavelength modes which differ
qualitatively in their form from those predicted
by the linear theory. However, including
nonlinearities, diffusion and noise, as we
remarked in the previous subsection, introduces
an enormous number of phenomenological
parameters.  We note
instead that the important new physics of
(\ref{eom}), namely, the wavelike (stable case)
or growing modes (unstable case), arises from 
the coupling of the vertical
and horizontal displacement fields, for
excitations with wavevector {\em transverse} to
the direction of mean drift, while the modes with
wavevector along $z$ play a relatively minor
role. This suggests that much can be learnt from
a model in {\em one} space dimension, the $x$
direction, corresponding to the $\perp$ direction
of (\ref{eom}), but retaining a {\em two}-component
displacement field ${\bf u} = (u_x, u_z)$. The
symmetry $x \rightarrow -x, \, u_x \rightarrow
-u_x$ then yields, to bilinear order in fields
and leading orders in gradients, the equations of
motion
\begin{mathletters}
\begin{eqnarray}
\dot{u_x}& = &{\lambda}_2 {{\partial}_x}u_z
+ {\gamma}_1 \partial_x u_x \partial_x u_z 
+ D_1 {{\partial}_x}^2 u_x +f_x
\label{eom1dconc}
\\
\dot{u_z} &=& {\lambda}_3 \partial_x u_x 
+ {\gamma}_2 (\partial_x u_x)^2 
\nonumber
\\
&&+ {\gamma}_3 (\partial_x u_z)^2 
+ D_2 {{\partial}_x}^2 u_z 
+f_z, 
\label{eom1dtilt}
\end{eqnarray}
\label{eom1d}
\end{mathletters}

\noindent which have, in addition to the $\{\lambda_i\}$, three nonlinear
coupling parameters $\{\gamma_i\}$ (also proportional to the driving
force $F$), two diffusivities $\{D_i\}$, and gaussian spatiotemporally
white noise sources $f_i, \, i = x,z$, with zero mean, and variances
$N_x, \, N_z$:
\begin{equation}
\langle f_i(0,0)f_j(x,t)\rangle = 2 N_i \delta_{ij}
\delta(x) \delta(t). 
\label{fcor}
\end{equation}
If $\{\gamma_i\}, \{D_i\}$ and $\{f_i\}$ are set to zero, we recover
the continuum limit of the equation derived by Crowley \cite{crowley}
for the dynamics of the small transverse and longitudinal displacements 
of a collection of hard spheres of radius $a$, prepared initially 
in a horizontal, one-dimensional periodic array with spacing $d$, 
settling vertically in a highly viscous 
fluid, with the hydrodynamic interaction cut off at the nearest 
neighbour scale.  The correspondence is 
$\lambda_2 = - \lambda_3 = - (3/4)a/d$, in units of the 
Stokes settling speed of an isolated sphere.  
Crowley's calculation can be extended beyond linear order to give
$\{\gamma_i\}$, but the elastic forces and the thermal fluctuations
that give the $D_i$s and $f_i$s are absent in his model.  The
diffusion and nonlinear terms in (\ref{eom1d}) are identical in
structure to those in the Erta\c{s}-Kardar (EK) models for the
fluctuations of drifting lines \cite{ertas1,ertas2}, with $u_x, \,
u_z$ replaced by their variables $h_{\perp}, \, h_{||}$ in
\cite{ertas1} or $R_{\perp}, \, R_{||}$ in \cite{ertas2}.  The EK
models, however, as a result of a larger symmetry ({\em independently}
under (i) $x \rightarrow -x$ and (ii) $R_{\perp}
\rightarrow - R_{\perp}$ or $h_{\perp}
\rightarrow - h_{\perp})$ lack
the linear first spatial derivative terms (the
$\lambda_i$ terms) of (\ref{eom1d}). 
Such linear terms can however be induced through 
the nonlinear terms, in \cite{ertas1,ertas2} by
constraining the ends of the line (polymer) to be at fixed mean
separation normal to the drift direction, so that
$\langle {\partial R_{\perp}
\over \partial x} \rangle \neq 0$. 
The related coupled-interface model of Barab\'asi \cite{barabasi}  
has an $x \rightarrow -x$ symmetry 
and thus also lacks the $\lambda_i$ terms of (\ref{eom1d}).  
These models are thus not relevant to 
the case of greatest interest to us here, namely,  
the unstable case $\lambda_2 \lambda_3 < 0$ of 
(\ref{eom1d}). 

In the unstable case, within a linear treatment, the concentration 
$\partial_x u_x$ and the tilt $\partial_x u_z$ grow without bound  
\cite{ashwin}. Physically, since real colloidal crystals 
are made of impenetrable particles, and since the elasticity of the
lattice will not tolerate arbitrarily large shear-strains, the
description implicit in (\ref{eom1d}) of small distortions about a
perfect lattice must break down in conditions of unstable growth. It
is best, therefore, to work from the outset with naturally bounded
variables for the concentration and tilt. To this end, we first pass
to a description in terms of the concentration fluctuation field
\begin{equation}
\sigma(x,t) = {\partial u_x \over \partial x}
\label{sigxt}
\end{equation}
and the tilt field  
\begin{equation}
\tau(x,t) = {\partial u_z \over \partial x}. 
\label{tauxt}
\end{equation}
Then (\ref{eom1d}) can be rewritten in the ``conservation-law'' 
form 
\begin{mathletters}
\begin{eqnarray}
\dot{\sigma}& = &{\lambda}_2 {{\partial}_x} \tau 
+ {\gamma}_1 \partial_x (\sigma \tau)  
+ D_1 {{\partial}_x}^2 \sigma + \partial_x f_x
\label{conccons}
\\
\dot{\tau} &=& {\lambda}_3 \partial_x \sigma  
+ {\gamma}_2 \partial_x (\sigma^2)
\nonumber
\\
&&+ {\gamma}_3 \partial_x (\tau ^2)
+ D_2 {{\partial}_x}^2 \tau  
+\partial_x f_z. 
\label{tiltcons}
\end{eqnarray}
\label{1dcons}
\end{mathletters}

\noindent As stated above, $\sigma$ and $\tau$ should be bounded;  
what matters on large length scales is only whether 
the local concentration is 
large or small compared to the mean, and whether the local 
tilt is ``up'' or ``down''. Accordingly,  
we construct a description in the next section in which the 
concentration and tilt fields of (\ref{1dcons}) are replaced 
by Ising variables evolving under a spin-exchange dynamics 
designed to mimic the most important aspects of (\ref{1dcons}). 
A continuum model 
which incorporates saturation is presented in section \ref{contsps}.  

\section{strong phase separation in a lattice model} 
\label{spslat}
In this section, we introduce the notion of Strong Phase Separation in
connection with the LR lattice model, which describes two coupled
species of spins on a lattice, with simple evolution rules which mimic
the coupled dynamics of the density and tilt fields.  This
coupled-spin problem is too difficult to solve for the dynamics or,  
indeed, for the steady state for arbitrary values of parameters.
However, for the symmetric case of half filling of both species, and a
special relation between coupling constants, we show 
(Section \ref{hamdetbal}) that
the condition of detailed balance is satisfied with respect to a
Hamiltonian ${\cal H}$ with long-ranged interactions.  In turn, this
allows for a characterization of the steady state of the system.  In
Section \ref{thermosps}, we show that at zero temperature $T$, 
the system exhibits
phase separation.  Moreover, we calculate thermodynamic properties and
show that the phase separation survives at {\it all} finite
temperatures, which is why we call this phenomenon Strong Phase
Separation (SPS).  The occurrence of SPS is linked to the long
(actually infinite) range of the interactions in ${\cal H}$, which
results in the energy being superextensive (proportional to $L^2$
rather than $L$).  We emphasise that this happens although the
underlying dynamical model is entirely local, with finite, bounded
rates. In section \ref{coarsening}, we show that this 
unusually robust phase separation sets in anomalously 
slowly, with domain sizes growing as the logarithm of 
time. The survival of SPS away from the detailed-balance limit 
is discussed through a kinetic interpretation in 
section \ref{nonsymcurrent}. 

\bigskip

\subsection{The LR Lattice Model}
\label{LRmodel}
\medskip

>From the study of driven diffusive systems, it is well known that
hydrodynamic behaviour can be recovered from the large-distance
long-time behaviour of simple lattice gas models evolving by
stochastic dynamics \cite{spohn}.  An example of such a model is the asymmetric
exclusion process, in which particles on a lattice perform biased
random walks subject to the constraint of no more than one particle
per site; in the limit of large separations and time, density
fluctuations are described by the Burgers equation with an additional
noise term.  An advantage of a lattice gas description is that
nonlinearities are incorporated implicitly in the nature of the
variable -- for instance, a $(0,1)$-valued occupation variable
incorporates the effects of exclusion.

Are there simple lattice gas models which capture the essential
features of coupled density-tilt dynamics of the type discussed in the
previous section?  Any such lattice model must, of course, involve two
sets of variables -- say $\{\sigma_i\}$ and $\{\tau_i\}$ -- which are
discrete versions of density and tilt fields and which evolve by rules
which mimic the physics of sedimenting lattices.  There are two
crucial features of the $\sigma - \tau$ dynamics of Eq. (\ref{1dcons}): 
first, that both $\sigma$ and $\tau$ fields are conserved so that their time
derivatives involve the divergences of currents; and second that the
local field which guides the $\sigma$-current has a term which is
proportional to $\tau$, and {\it vice versa}.  Accordingly, we 
define \cite{rlsr} 
a lattice model which incorporates just these effects.  Consider a
one-dimensional lattice made of two interpenetrating sublattices $S$
($i = 1,2,3,\cdots,N$) and $T$ ($i =
{1\over2},{3\over2},\cdots,N-{1\over2}$).  Place Ising variables
$\sigma_i = \pm 1$ at every site of $S$, and $\tau_{i+{1\over2}} = \pm
1$ on every site of $T$.  We take $\sigma_i = 1$ if there is a
particle at site $i$, and $\sigma_i = -1$ if there is no particle,
while $\tau_i = 1$ or $-1$ denotes the two possible values of the
local tilt.  The dynamics involves exchange of adjacent spins
$\sigma_i$ and $\sigma_{i+1}$ at a rate which depends on the
intervening spin $\tau_{i+{1\over2}}$, while the rate of $\tau$-spin
exchanges depends on the intervening $\sigma$ spin, i.e. we have
Kawasaki \cite{kawa} dynamics, with hopping rates which depend on the 
local value of the other species.  The probability $P(C)$ that the system is
in a configuration $C \equiv (\{\sigma_i\},\{\tau_{i-{1\over2}}\})$
evolves through the master equation
\be
{dP(C) \over dt} = \sum_{\langle n,n+1\rangle} W(C_{n,n+1} \rightarrow C)
P(C_{n,n+1}) - W(C \rightarrow C_{n,n+1}) P(C).
\label{master}
\ee
Here $\langle n,n+1\rangle$ on the right hand side (with $n =
{1\over2},1,{3\over2},2,\cdots$) labels transitions which involve pairwise
interchanges of neighbouring $\sigma$'s $(\sigma_i \leftrightarrow
\sigma_{i+1})$ and $\tau$'s $(\tau_{i-{1\over2}} \leftrightarrow
\tau_{i+{1\over2}})$, and configuration $C_{n,n+1}$ differs from $C$
only through the interchange of spins on site $n$ and $n+1$.  The most
general such model would involve the 8 distinct transition rates
listed in Eq. (1).  For a left-right symmetric system, we have 
$D = D'$; $a = a'$; $sgn(b) = sgn(b')$: this
defines the LR model \cite{rlsr}.

In the interest of defining a {\em minimal version} of the LR model, we also
impose the further restrictions $E = E'$, $b = b'$.  The rates of
the minimal model may be written compactly as
\[
W (\sigma_i \leftrightarrow \sigma_{i+1}; \tau_{i+{1\over2}}) = D
- {a \tau_{i+{1\over2}} \over 2} (\sigma_i - \sigma_{i+1})
\]
\be
W(\tau_{i-{1\over2}} \leftrightarrow \tau_{i+{1\over2}}; \sigma_i) = E
+ {b \sigma_i \over 2} (\tau_{i-{1\over2}} - \tau_{i+{1\over2}}).
\label{minrates}
\ee
The evolution rules can be stated as follows: If $a$ is positive, a
particle tends to move downhill, and a hole uphill.  If $b$ is
positive, a local peak $(\wedge)$ tends to transform into a valley
$(\vee)$ if a particle
resides on it, while local valleys tend to become peaks in the
presence of holes.  Changing the signs of $a$ and
$b$ reverses these tendencies.  As a result, the nature of the
steady state is sensitive to the sign of $\alpha \equiv ab$.  As we
will see below, if $\alpha$ is positive, the exchanges of $\sigma$ and
$\tau$ spins in Eq. (\ref{minrates}) act in concert to promote segregation of
both species of spins, ultimately resulting in a phase-separated
state.  This is the unstable case of the LR model -- the case of
primary interest in this paper.  By contrast if $\alpha$ is negative,
`easy' $\sigma$ and $\tau$ moves produce opposing tendencies, and
hence result in a fluctuating but on-average spatially homogeneous state -- the
stable case of the LR model. The calculations of Crowley \cite{crowley} 
for settling arrays of hydrodynamically interacting spheres and the discussion 
in \cite{rlsr} make it clear that for sedimenting colloidal crystals 
it is the ``unstable'' case that applies. 

The other important parameters in the model are the magnetizations $M_\sigma 
\equiv \Sigma_i \sigma_i/N$, $M_\tau = \Sigma_i \tau_{i+{1\over2}}/N$,
both of which are conserved by the dynamics. 
\bigskip

\subsection{Symmetric Case: Hamiltonian and Detailed Balance}
\label{hamdetbal}
\medskip

We now consider the {\em symmetric case} of the LR model, which is defined
by the vanishing of the magnetizations
\be
M_\sigma = M_\tau = 0,
\label{zerotilt}
\ee
and the following relationship between coupling constants in (\ref{minrates})
\be
{b \over E} = {a \over D}.
\label{ratios}
\ee
Since $E, \, D > 0$, it is clear that (\ref{ratios}) is a special case
of the {\em unstable} LR model.  We show below that when conditions
(\ref{zerotilt}) and (\ref{ratios}) are met, it is possible to
find a Hamiltonian ${\cal H}$ such that the condition of detailed
balance is satisfied with invariant measure $\exp(-\beta {\cal H})$.

\begin{figure}
\epsfxsize=12cm
\centerline{\epsfbox{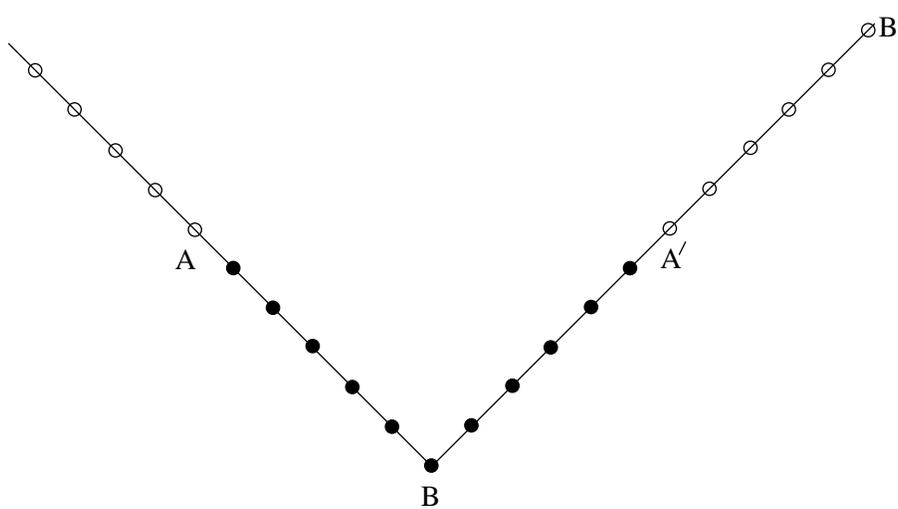}}
\caption[]{\label{fullphasesep}The phase-separated state of the 1-d lattice model at T=0
is shown. $\sigma$ and $\tau$ variables are shown as circles and squares
respectively, with $\sigma, \tau=+1(-1)$ shown filled (empty).
The configuration of the corresponding height model is also shown.
Interfaces between $\sigma=+1$ and $-1$ are located at $A$ and $A'$, and
those between $\tau=+1$ and $-1$ are at $B$ and $B'$.}
\end{figure}

Since the motion of $\sigma$ particles is determined by the local tilt
$\tau$, we may think of the $\sigma$ particles as moving in a
potential landscape provided by the $\tau$'s (Fig. \ref{fullphasesep}).  With this in
mind, we define the height at site $k$ by
\be
h_k \{\tau\} = \sum^k_{j=1} \tau_{j-1/2}.
\label{fieldh}
\ee
With
periodic boundary conditions $(\sigma_{N+i} = \sigma_i; \
\tau_{N+i-{1\over2}} = \tau_{i-{1\over2}})$, the zero-net-tilt
condition $M_\tau = 0$ implies 
$h_{N+k} = h_k$.  We associate a potential energy proportional to $h_k
\sigma_k$ with site $k$, and write the Hamiltonian
\be
{\cal H} = \epsilon \sum^N_{k=1} h_k \{\tau\} \sigma_k
\label{hamh}
\ee
to describe the total energy of the $\sigma$ particles in the
landscape derived from the $\tau$ particles.

In view of the symmetric role played by $\sigma$'s and $\tau$'s in the
Symmetric Model, we may equally ask for the potential energy of $\tau$
particles in the landscape provided by the $\sigma$ particles.  The
corresponding Hamiltonian is then
\be
{\cal G} = \epsilon \sum^N_{k=1} g_{k-{1\over2}} \{\sigma\}
\tau_{k-{1\over2}}
\label{hamg}
\ee
where the height $g$ is given by
\be
g_{k+{1\over2}} \{\sigma\} = \sum^k_{j=1} \sigma_j.
\label{fieldg}
\ee
When the heights $h_k$ and $g_{k-{1\over2}}$ are written out in terms
of $\tau_j$'s and $\sigma_j$'s respectively, the Hamiltonians ${\cal
H}$ and ${\cal G}$ are seen to involve very nonlocal couplings:
\be
{\cal H} = \epsilon \sum^N_{k=1} \sum^k_{j=1} \tau_{j-{1\over2}} \sigma_k
\label{hamh2}
\ee
\be
{\cal G} = \epsilon \sum^N_{k=2} \sum^{k-1}_{j=1} \sigma_j
\tau_{k-{1\over2}}.
\label{hamg2}
\ee
We observe that
\be
{\cal H} + {\cal G} = \epsilon M_\sigma M_\tau 
\label{hamrel}
\ee
and, since each of $M_\sigma$ and $M_\tau$ vanishes in the symmetric
case owing to the zero-tilt condition, we have ${\cal H} = -{\cal G}$.
Thus the Hamiltonians corresponding to the two pictures i.e. $\sigma$
particles in a $\tau$-landscape or {\it vice versa} are completely
equivalent.  We will mostly use ${\cal H}$ for further work.

We now show that the steady state of the Symmetric model defined by
Eqs. (\ref{zerotilt}) and (\ref{ratios}) satisfies the condition of
detailed balance and that the stationary measure is given by
$e^{-\beta {\cal H}}$ where $\beta$ is the inverse temperature
$T^{-1}$, with $\beta \epsilon$ given by (\ref{b1}) below.  To this
end, let us ask for the changes in energy $\Delta E (\sigma_i
\leftrightarrow
\sigma_{i+1})$ of ${\cal H}$ when spins $\sigma_i$ and $\sigma_{i+1}$
are interchanged, and $\Delta E(\tau_{i-1/2} \leftrightarrow
\tau_{i+1/2})$ in ${\cal H}$ when spins $\tau_{i-{1\over2}}$ and
$\tau_{i+1/2}$ are interchanged.  For $i \neq N$, it is
straightforward to see that 
\be
\Delta E(\sigma_i \leftrightarrow \sigma_{i+1}) = \epsilon
\tau_{i+{1\over2}} (\sigma_i - \sigma_{i+1})
\label{change1}
\ee
\be
\Delta E(\tau_{i-{1\over2}} \leftrightarrow \tau_{i+{1\over2}}) =
\epsilon \sigma_i
(\tau_{i+{1\over2}} - \tau_{i-{1\over2}})
\label{change2}
\ee                                                                   
In fact, Eqs. (\ref{change1}) and (\ref{change2}) are valid for $i=N$
as well, as can be verified on 
recalling that $\sigma_{N+1} = \sigma_1$, $\tau_{N+{1\over2}} =
\tau_{1/2}$ and using the zero-tilt conditions $M_\sigma = M_\tau = 0$
while computing energy changes. 

Consider the configuration $C_{\sigma_i,\sigma_{i+1}}$ obtained from a
configuration $C$ on exchanging two neighbouring $\sigma$ spins -- an
elementary kinetic move in the model.  The condition of
detailed balance is then
\be
{W(C \rightarrow C_{\sigma_i,\sigma_{i+1}}) \over
W(C_{\sigma_i,\sigma_{i+1}} \rightarrow C)} = {\mu_{SS}
(C_{\sigma_i,\sigma_{i+1}}) \over \mu_{SS} (C)}
\label{detbal}
\ee
where $\mu_{SS}(C)$ is the steady-state measure for configuration
$C$.  To verify that
\be
\mu_{SS} (C) = e^{-\beta {\cal H} (C)},
\label{meas}
\ee
we use Eqs. (\ref{minrates}) and (\ref{change1}) to obtain
\be
{D - aX_i \over D + aX_i} = e^{-2\beta \epsilon X_i}
\label{bal}
\ee
where we have defined $X_i \equiv {1\over2} \tau_{i+{1\over2}}
(\sigma_i - \sigma_{i+1})$.
Noting that $X_i = \pm1$, we see that Eq. (\ref{bal}) is satisfied provided
\be
\beta\epsilon = {1\over2} \ell n\left({D+a \over D-a}\right).
\label{b1}
\ee
In order for the measure to be valid under interchanges of adjacent
$\tau$'s $(\tau_{i-{1\over2}} \leftrightarrow \tau_{i+{1\over2}})$,
similar reasoning leads to the condition
\be
\beta\epsilon = {1\over2} \ell n\left({E+b \over E-b}\right).
\label{b2}
\ee
In the Symmetric case of the LR Model, Eq. (\ref{ratios}) holds, and so
Eqs. (\ref{b1}) and (\ref{b2}) are consistent.  Thus 
the condition of detailed balance holds with the equilibrium measure
(\ref{meas}). 

It is appropriate to recall that the three-species model of Evans et
al. \cite{c3cm} also obeys the condition of detailed balance in the
symmetric case. There too the Hamiltonian has infinite ranged
interactions, but does not have as transparent an interpretation as
(\ref{hamh}). 

\bigskip

\subsection{Symmetric Case: Thermodynamic Properties and Strong
Phase Separation} 
\label{thermosps}
\medskip

Since the condition of detailed balance holds in the symmetric case of
the minimal LR model, the steady state corresponds to the thermal equilibrium
state with Hamiltonian ${\cal H}$.  The thermodynamic properties of
the system can be found, in principle, using equilibrium statistical
mechanics.  A calculation can be carried out in the grand canonical ensemble
in the limit $N \rightarrow \infty$.  The
resulting state exhibits Strong Phase Separation. 

The Hamiltonian ${\cal H}$ (Eq. (\ref{hamh})) describes spins
$\sigma_k$ in a site-dependent magnetic field $\epsilon h_k$, which is
itself a dynamical variable.  Equivalently, in the lattice gas
description (associating an occupation variable $n_k = {1\over2} (1 +
\sigma_k)$), it describes particles with a hard core constraint in a
potential well of depth $\epsilon h_k$.  The ground state of ${\cal
H}$ is obtained by arranging the $\tau$ spins (which determine the
heights $h_k$) so as to form as deep a potential well as possible, and 
then arranging the $\sigma$-particles at the bottom of the well
(Fig. \ref{fullphasesep}).  A spin configuration which corresponds to this choice is
\beq 
\tau_{k-1/2} &=& -1 \ {\rm for} \ k = 1,\cdots,N/2 \nonumber \\[2mm]
&=& 1 \ {\rm for} \ k = {N \over 2} + 1,\cdots,N \nonumber \\[2mm]
\sigma_k  &=& 1 \ {\rm for} \ k = N/4,\cdots,3N/4 \nonumber
\\[2mm] &=& -1 \ {\rm for} \ k = 1,\cdots,N/4-1 \ {\rm and} \ k =
3N/4+1,\cdots,N. \label{grdst} 
\eeq 
Each spin species exhibits
complete phase separation in this ground state.    The ground state
energy is straightforward to compute, 
and we find \be E_G \simeq - {\epsilon N^2 \over 8}.
\label{grden}
\ee
Notice the quadratic dependence of $E_G$ on $N$, which is an outcome
of the infinite-ranged interactions in ${\cal H}$ (Eq. \ref{grden}).  
As explained below, this
unusual superextensive behaviour of the energy is ultimately the
feature responsible for the phenomenon of Strong Phase Separation,
namely the continued existence and stability of the phase separated
state at all finite temperatures.

\begin{figure}
\epsfxsize=12cm
\centerline{\epsfbox{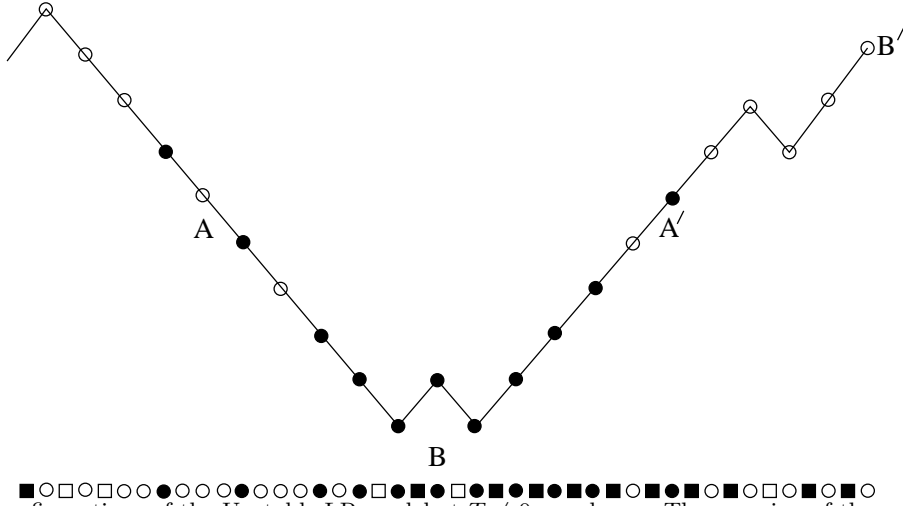}}
\caption[]{\label{partphasesep} Typical configurations of 
the Unstable LR model at $T \ne 0$ are shown. The meaning of the symbols 
is as in Fig. \ref{fullphasesep}. 
Phase separation persists, but there are particle-hole excitations
of both species near the corresponding interfaces.} 
\end{figure}

At $T = 0$, phase separation is complete and there is a sharp boundary
between regions of positive and negative spins of each species.  Let
$A$ and $A'$ be the locations of the $T=0$ interface between regions
with $\sigma=1$ and $\sigma=-1$, and let $B$ and $B'$ be the locations
of interfaces separating regions with $\tau =1$ and $\tau = -1$
(Fig. \ref{fullphasesep}).  The effect of raising the temperature to a
finite value $T$ is to smear out the interfacial zones around $A$,
$A'$, $B$ and $B'$ (Fig. \ref{partphasesep}).  To address this
quantitatively, let us turn to the evaluation of thermodynamic
properties.

The calculation can be carried out most easily in a grand ensemble in
which the total magnetizations $M_\sigma$ and $M_\tau$ are not held
fixed. The corresponding grand partition function is 
\beq
Z &\equiv& \sum_{\{\sigma\},\{\tau\}} e^{-\beta({\cal H} - E_G)}
\nonumber \\[2mm] &=& \sum_{\{\sigma\}\{\tau\}} e^{-\beta\epsilon
\sum_k h_k (\sigma_k - \sigma^0_k)} \label{z1}
\eeq
where $\sigma^0_k$ denotes the value of $\sigma_k$ in the ground
state.  The key observation that allows the calculation to be
performed is that near
the $\sigma$ interfaces $A$ and $A'$, the field $h_k(\{\tau\})$ is
essentially fixed at its $T=0$ value $h^0_k$; deviations are of order
$\exp[-\beta\epsilon N/4]$ as explained below, and so are utterly
negligible in the thermodynamic limit.  Likewise, in the vicinity of
the $\tau$-interfaces $B$ and $B'$, the $\sigma$ spins are frozen to
their $T=0$ values, and so $g_k(\{\sigma\}) = g^0_k$.  To proceed,
let us divide
the system into four equal parts $R_A$, $R_B$,
$R_{A'}$, $R_{B'}$, where region $R_A$ consists of the $N/4$ spins of
each of the two species centered around $A$.  Other regions are defined
similarly, centered around $B$, $A'$ and $B'$.  Evidently, with
negligible error we may set $h_k(\{\tau\}) = h^0_k$ in regions $R_A$
and $R_{A'}$, and set $g_k(\{\sigma\}) = g^0_k$ in regions $R_B$ and
$R_{B'}$.  The partition function $Z$ can then be written as the
product of 4 terms $Z_A$, $Z_B$, $Z_{A'}$, $Z_{B'}$, where for
instance
\beq
Z_A &=& \sum_{\{\sigma\}} e^{-\beta\epsilon \sum_{k\in R_A}
h^0_k(\sigma_k - \sigma^0_k)} \label{za1} \\[2mm] Z_B &=& \sum_{\{\tau\}}
e^{-\beta\epsilon \sum_{k\in R_B} g^0_{k+{1\over2}}(\tau_{k+{1\over2}} -
\tau^0_{k+{1\over2}})}. \label{zb1}
\eeq
Each of these factorizes into single-site partition functions, and can
be evaluated straightforwardly.  Recalling that $h^0_k$ varies
linearly with $k$ near the $T=0$ interface location $k_A$, we find 
\be
Z_A = \prod_{k\in R_A} (1 + e^{-2\beta\epsilon|k - k_A|}).
\label{za2}
\ee
In the thermodynamic limit, we obtain
\be
Z_A = K(e^{-2\beta\epsilon})
\label{za3}
\ee
where $K(y) \equiv \prod^\infty_{k=-\infty} (1 + y^{|k|})$ is a generating
function that arises in the theory of partitions \cite{hardy}.
Evidently, each of $Z_B$, $Z_{A'}$ and $Z_{B'}$ equals the same
quantity as well, so that $Z=\left[K(\exp(-2\beta\epsilon))\right]^4$.

It is worth pausing to comment on the unusual size dependences of
various quantities.  The ground state energy $E_G$ is proportional to
$N^2$, a superextensive dependence.  This has its origin in the
infinite-ranged interactions in $\cal{H}$.  Further, with energies
measured from the ground state value, the partition function
approaches an $N$-independent limit.  Thus the total change in free
energy and entropy away from $T=0$ remain finite in the thermodynamic
limit {\it i.e.} they are not extensive.  This reflects the fact that
the only effect of raising the temperature is to broaden the
interfacial region between phases, which essentially affects only a
finite number of sites.

In fact, an explicit calculation of the broadened interfacial profile
can be carried out in the grand ensemble.  For instance, near $A$ we have
\be
\langle \sigma_k \rangle = \tanh \beta \epsilon h^0_k
\label{profile}
\ee
where $h_k = (k-k_A)$.  We see
that $\langle \sigma_k \rangle$ deviates substantially from 1 only in
a region where $|\beta \epsilon h_k| \lsim 1$, or 
\be
|k - k_A| < T/\epsilon.
\label{width2}
\ee
For sites $k$ such that
$|k-k_A| \gg T/\epsilon$, the deviation from $\pm 1$ is $\approx
2\exp(-2\beta\epsilon|k-k_A|)$ which vanishes rapidly.  We see that the primary
effect of temperature is to smear out the interfaces.  The formation
of `droplets' far from the interfaces is prohibitively costly in
energy, and hence the probability dies down exponentially.  Recalling that the
separation of the two $\sigma = 1 \rightarrow -1$ interfaces is $N/2$,
in the thermodynamic limit $N \rightarrow \infty$, we see that only a vanishing
fraction of spins (those close to the interfaces) deviate from values
arbitrarily close to 1 and $-1$.  In this sense, phase separation
remains complete and cannot be effaced at any finite temperature $T$,
i.e. we have Strong Phase Separation.

These results obtained in the grand ensemble provide a qualitative, if 
not quantitative, guide to the thermodynamic properties of the system
in which $M_\sigma$ and $M_\tau$ are held fixed. The customary
equivalence between ensembles is not obviously valid any longer, as
particle-hole excitations are essentially confined to a finite
region of width proportional to $T$, which does not increase as $N \rightarrow \infty$.
Thus the difference between observables
calculated in the two ensembles is expected to remain of order unity,
and not die out in the $N \rightarrow \infty$ limit \cite{kavita}. Interestingly, the
calculation of the partition function, though not the profile,
has been carried out for the three-species model within a 
constant-species-number ensemble \cite{c3cm}.

The stability of the strongly phase-separated state can also be
understood in terms of kinetics.  In the ground state arrangement of
Fig. \ref{fullphasesep}, each $\sigma$ spin finds itself in a uniform field produced by
the $\tau$ spins.  Consider moving a spin over a macroscopic distance
-- say a $\sigma = +1$ spin from $A'$ to $A$, via $B$.  The movement
from $A'$ to $B$ may be viewed as an activation process as the spin in
question has to overcome a potential barrier of magnitude $\epsilon
N/4$ to reach $B$; beyond that, in the region $BA$, the motion is
ballistic as the $\tau$-induced field helps it along.  The
rate-limiting step is thus the $A' \rightarrow B$ activation.  At
temperature $T$, the relevant time scale is of the order of $t_{CB}
\sim \exp(\epsilon N/4T)$ which diverges rapidly as $N \rightarrow
\infty$.  Thus, in the thermodynamic limit, a rearrangement of the SPS
state is not possible; the only effect of the temperature-assisted
motion is to move a few $\sigma = 1$ spins near the interface into the
$\sigma = -1$ rich region and vice versa, but such penetration does
not proceed far in view of the restoring fields.  Defining the
penetration depth $\Delta k$ as that over which the activation time
falls by a factor of $1/e$, we estimate $\Delta k = T/\epsilon$, in
agreement with Eq. (\ref{width2}) which was based on the spatial decay 
of the interfacial profile.

\subsection{Coarsening}
\label{coarsening}
Now imagine that Fig. \ref{fullphasesep} represented {\em half} 
the system, and that the other half was identical in structure. 
This would amount to a system that had phase-separated into {\em four} 
macroscopic domains, each of size $N/8$. For this state to proceed towards 
full phase separation, the two $+$ domains, each at the bottom 
of a valley, must merge. The rate-limiting step can again be taken to be 
the movement of a $+$ from the edge of an all\,$-+$ region 
to the top of a hill, i.e., a distance $N/8$. Once this comes to pass, 
the two domains of length $N/4$ will rapidly merge to give one domain 
of length $N/2$. The time for this, which is the time for 
complete phase separation for a system of size $N$, can be seen 
from the argument in section \ref{thermosps} to scale as 
exp$(\epsilon N / 8T)$. This tells us that the characteristic 
domain size grows logarithmically in time, as stated in section \ref{intro}. 
%Despite this ultraslow growth, 
%the time [$\sim$ exp$(N/4T)$] required for the macroscopic remixing 
%of a fully phase-separated state is overwhelmingly larger than that 
%[$\sim$ exp$(N/8T)$] for domains to grow to a size $N$, reinforcing 
%our conclusion in section \ref{thermosps} that the system is 
%macroscopically segregated at thermal equilibrium.  

The time required for the reverse process (from a 2-domain to a
4-domain state) scales as $\sim \exp [(N/4T)]$, which is
overwhelmingly larger than the $4 \rightarrow 2$ coarsening time.
This is true at all scales, and the transition from a $2n$-domain
state to one with $n$ domains is much more rapid than the reverse.
Thus the transition from a statistically homogeneously mixed state to
the equilibrium phase-separated state is irreversible, even though it
occurs slowly.

The coarsening process was studied \cite{c3cm} both numerically in the 3-species
model and within a mean-field approximation for a related `toy' model.
The typical domain size was found to grow logarithmically 
in time. The arguments given above are consistent with this.

\bigskip

\subsection{Non-symmetric Case}
\label{nonsymcurrent}
\medskip

We now address the nature of the steady state for arbitrary values of
$M_\sigma$ and $M_\tau$.  Away from $M_\sigma = M_\tau = 0$ the
problem is no longer described in terms of the equilibrium state of a
long-ranged Hamiltonian; nevertheless we will argue below that the
system continues to exhibit Strong Phase Separation.

\begin{figure}
\epsfxsize=12cm
\centerline{\epsfbox{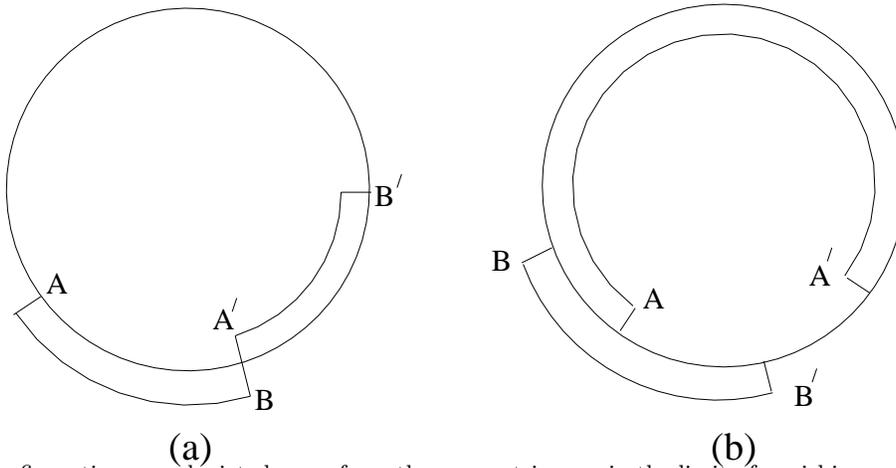}}
\caption[]{\label{interfaces}Typical configurations are depicted away from the symmetric case
in the limit of vanishing noise. (a) If the fraction of $\sigma =+1$
and $\tau=+1$ spins is low enough, interfaces $A'$ and $B$ coincide.
(b) If the fraction of $\sigma=+1$ spins is high enough, the interface
$B'$ lies halfway between $A$ and $A'$.}
\end{figure}

It is useful to define $x$ and $y$ as the density of up spins of the
$\sigma$ and $\tau$ types.  We have $x = {1\over2} (1 + M_\sigma/N)$
and $y = {1\over2} (1 + M_\tau/N)$. 
If $x$ and $y$ are small enough that $2x+y<1$, the steady state 
is of the type shown
in Fig. \ref{interfaces}a, with each of the $\sigma$ and $\tau$ species showing phase
separation, but with basically no spatial overlap of the $\sigma = 1$
and $\tau = 1$ regions.  A useful way to characterize this state is
through the sequence of interfaces, viz. $A \cdots B' \cdots BA'
\cdots$, where $\cdots$ denotes a macroscopic stretch of the system.
Here $A(B)$ separates an up-spin region of $\sigma(\tau)$ spins on the
right, from the corresponding down-spin regions, while $A'$ and $B'$
separate the opposite regions.  Trial states of the type $A \cdots B'
\cdots B \cdots A' \cdots$ are seen to approach the non-overlapping
state on a time scale of order $t^\star$ where $\ln t^\star$ is
less than but of the order of the smaller of $\epsilon N x/T$ and
$\epsilon N y/T$.  Once the nonoverlapping steady state has been
reached, $\sigma$ and $\tau$ spins can still be cycled around by
activation processes across $A'A$ and $A'A$ respectively (Fig. \ref{interfaces}), but
such cycling around does not change the character of the state.

Now consider increasing $y$, keeping $x$ fixed.  The number of spins
in the stretch between $B'$ and $A$ is $N(1 - x - y)$, and once this
drops below $Nx$, the predominant activation process occurs over this
stretch.  Thus the no-overlap state of Fig. \ref{interfaces}a is unstable towards a state
of the type shown in Fig. \ref{interfaces}b, once $2x+y$ exceeds unity.
In this steady state, activation
processes in a finite system lead to small currents of $\sigma$ and
$\tau$ spins, of magnitude
\be
\label{Sigmacurr}
J_\sigma = a_1 \exp(-\epsilon \ell_{AB'}/T) - a_2 \exp(-\epsilon
\ell_{A'B'}/T) 
\ee
\be
\label{Taucurr}
J_\tau = a_3 \exp(-\epsilon \ell_{BA}/T) - a_4 \exp(-\epsilon
\ell_{B'A}/T) 
\ee
where $a_1,a_2,a_3,a_4$ are prefactors of order unity and $\ell_{AB'}$
is the separation of interfaces $A$ and $B'$, and other $\ell$'s are
defined similarly. Since the difference $\ell_{A'B'} - \ell_{AB'}$ is 
positive and grows
proportionally to $N$, we may drop the second term on the right of
Equation (\ref{Sigmacurr}).  In steady state we must have $J_\sigma =
J_\tau$, which then leads to
$(a_1 + a_4) \exp(-\epsilon \ell_{AB'}/T) \simeq a_3 \exp(-\epsilon
\ell_{BA}/T) $
or
\be
\label{Cond2}
\ell_{AB'} = \ell_{BA} + {\rm terms \ of \ order \ unity}.
\ee
Thus, $A$ is very close to the halfway position in stretch $BB'$. The
overlapping stretch $BA'$ is a fraction 
$\delta = {1\over2} (2x + y - 1)$ of the whole length.
On setting $x = y = {1\over2}$, we recover $\delta = 1/4$ in agreement
with the results of the equilibrium analysis of Section C.

Analogously, keeping $y$ fixed and increasing $x$ we conclude that for
$2y + x > 1$, the steady state is $A \cdots B \cdots A' \cdots B'
\cdots$ with $B'$ between $A'$ and $A$ and an overlapping stretch of
length ${1\over2} N (2y + x - 1)$.  Finally, under $x \rightarrow
(1 - x)$, $y \rightarrow (1 - y)$ we arrive at the condition for
overlap of negative spins.  

In short, strong phase separation persists even away from the
symmetric point of the LR model. In general, two types of steady
states, both phase separated, are possible as depicted in Fig. \ref{interfaces}. In the overlapping case,
there is generally a current in a finite system, but this vanishes
exponentially with system size.
While we have explored the effects of deviating from the
symmetric case by moving away from the half-filling condition
Eq. (\ref{zerotilt}), without altering the condition (\ref{ratios}) on
the rates, another way to make the system nonsymmetric is to violate
the latter condition.  We have not explored this in detail but expect that
the phenomenon of SPS will persist in this case too so long as 
$ab > 0$.

%\newpage

\section{detailed balance and strong phase separation in 
the continuum limit} 
\label{contsps}
The continuum model of section \ref{contmodel}, in the 
case $\lambda_2 \lambda_3 < 0$ in equation 
(\ref{eom1d}), is linearly unstable. One way to 
deal with this instability is to resort, as we 
have done above, to a lattice model in which the 
variables are naturally bounded. An alternative 
way is to ask what nonlinear terms added 
to (\ref{eom1d}) for $\lambda_2 \lambda_3 < 0$ 
would arrest the unstable growth \cite{ashwin}.   
To do this, we work in the detailed-balance 
limit of the lattice model, start with the 
Hamiltonian (\ref{hamh}), and construct the 
corresponding continuum Ginzburg-Landau  
free-energy functional. We shall see 
below that this functional will give rise 
to dynamical equations with the same linear 
instability as in (\ref{eom1d}) with  
$\lambda_2 \lambda_3 < 0$, but containing 
nonlinearities which prevent unbounded growth.  
 
The derivation is straightforward, as  
the condition of detailed balance allows us to 
proceed as in any equilibrium statistical 
mechanics problem. The Ginzburg-Landau free-energy 
functional $F[\sigma, \tau]$ for our system, i.e., 
the effective Hamiltonian for 
a description in terms of the coarse-grained fields 
$\{\sigma(x), \, \tau(x) \}$ of section \ref{cont1d}, may be written as 
$U \, - \, T S$ where $U$ is the energy 
(\ref{hamh}) in the continuum limit,  
$T$ the temperature, and $S$ the entropy 
obtained by summing over all microscopic configurations 
$\{\sigma_i, \tau_i \}$ subject to a fixed coarse-grained configuration 
$\{\sigma(x), \, \tau(x) \}$. Since $\sigma_i$ and $\tau_i$ 
are Ising variables, $S$ can be found from a standard 
Bragg-Williams construction.  
Thus,  
\begin{eqnarray}
F\left[\sigma,\tau \right] =&& \epsilon \int_{0}^{L} \mbox{d}x 
\int_{0}^{x} \mbox{d}x' \sigma(x) \tau(x')\nonumber \\ 
+&& T \int_{0}^{L} \mbox{d}x \sum_{m = \tau(x), \sigma(x)}
\left[ {{1+m} \over 2} \ln{{1+m} \over 2} + {{1-m} \over 2} 
\ln{{1-m} \over 2}\right], 
\label{freeen}
\end{eqnarray}
where $x$ is measured in units of the lattice spacing 
and is hence dimensionless. The partition function is 
then $\int [\mbox{d}\sigma][\mbox{d}\tau] \mbox{exp}(-F/T)$.  
 
Then the usual, purely dissipative, {\em conserving} time-dependent
Ginzburg Landau equations of motion generated by (\ref{freeen}), {\em
i.e.},
\begin{equation}
\label{tdgl}
\partial_t \sigma = \Lambda_{\sigma} \partial^2_x {{\delta F} \over {\delta \sigma}} + \eta_{\sigma} 
\end{equation}  
and likewise for $\tau$, turn out to be precisely 
\begin{eqnarray}
\label{consnonloctdgl}
\partial_t \sigma &=& \Lambda_{\sigma} (T \partial^2_x 
\mbox{tanh}^{-1} \sigma + \epsilon \partial_x \tau) + \eta_{\sigma};  
\nonumber 
\\
\partial_t \tau &=& \Lambda_{\tau}(T \partial^2_x 
\mbox{tanh}^{-1} \tau - \epsilon \partial_x \sigma) + \eta_{\tau}.  
\end{eqnarray}
Here $\Lambda_{\sigma}, \, \Lambda_{\tau}$ are mobilities and 
$\eta_{\sigma}, \, \eta_{\tau}$ are noise sources with 
variances proportional to the corresponding mobilities.  
It is evident that equations (\ref{consnonloctdgl}) and 
(\ref{1dcons}) are identical in the linearised limit, 
if we make the identification $\lambda_2 = \Lambda_{\sigma} \epsilon, 
\, \lambda_3 = -\Lambda_{\tau} \epsilon$. This corresponds to the 
linearly unstable limit of (\ref{eom1d}), in consonance with 
fact that the detailed balance limit of the lattice model 
was derived in precisely that case.  

We should thus be able to gain some insight into SPS 
by looking at the steady states of (\ref{consnonloctdgl}). 
The simplest of these are the zero current states, which 
satisfy  
\begin{eqnarray}
\label{zerocurr}
\partial_x P - \mbox{tanh}Q &=& 0, 
\nonumber 
\\
\partial_x Q + \mbox{tanh}P &=& 0,    
\end{eqnarray}
where 
\begin{equation}
\label{pq}
Q \equiv \mbox{tanh}^{-1}\phi, \, P \equiv \mbox{tanh}^{-1}\psi.   
\end{equation}
The {\em spatial} development of $P$ and $Q$ with respect to $x$ is like 
a Hamiltonian dynamics, conserving the ``energy'' 
\begin{equation}
\label{pseuden}
E(P,Q) = \ln (\cosh P \cosh Q)
\end{equation}
This leads to closed orbits in the $P-Q$ or $\psi-\phi$ plane, i.e., 
regions of large $\psi$ and small $\phi$ followed by the opposite. 
These are spatially multidomain states which will not evolve further 
in the absence of noise.  
 
\section{Summary and Discussion}
\label{conclusion}
\subsection{Summary}
\label{summary} 
In summary, we have constructed continuum and lattice models
to describe the physics of steadily sedimenting colloidal crystals or, more
generally, of a crystal driven through a dissipative medium. The models
display two broadly distinct types of behaviour, termed ``stable''
and ``unstable'', depending on the sign of a parameter.  We have
concentrated on the unstable case and shown, through a mapping to a
one-dimensional lattice model, that it {\em always} displays phase
separation, a
phenomenon which we call strong phase separation. This phase
separation and the fact that it persists at all temperatures can be
understood, in general, in terms of barriers to remixing which grow
with system size.  The barriers are erected by the system in the
course of its evolution, and result in domain sizes growing as the
logarithm of the time. In a particular limit, the detailed balance
condition holds, allowing us to write the steady state distribution
in the equilibrium form $\exp (-\beta {\cal H})$, and to calculate
density profiles exactly.  Here $\cal H$ 
involves long ranged
interactions even though the model has strictly local dynamics.  This
long-ranged character of interactions in ${\cal H}$ is responsible 
for the phase
separation in this one-dimensional system, and the fact that it
persists at all temperatures. 

\subsection{Experimental tests}
\label{expts}
Finally, let us turn to the possibility of testing our results in
experiments.  We have demonstrated Strong Phase Separation in a {\em
one-dimensional} model system. It seems highly likely, therefore, that
the same phenomenon will take place in the experimental systems which
inspired our model, namely, steadily sedimenting crystalline
suspensions in, for example, the two-dimensional geometry described in
section \ref{intro}.  A good candidate system is a charge-stabilised
crystalline array of polystyrene spheres with radius in the micron
range. The lattice spacing of the crystal should be neither so large
that hydrodynamic effects (proportional to the ratio of particle size
to interparticle spacing) are negligible, nor so small that the flow is
choked. This will ensure that appreciable hydrodynamic flow takes
place between the spheres, giving rise to the strain-dependent
mobilities \cite{crowley} that are used in (\ref{eom}). If the system
parameters are as in \cite{rutgers}, the Reynolds number will be
negligible, as required by our neglect of inertia, and the Peclet
number large. Note that our model equations (\ref{eom}) were
formulated to describe the nature of distortions about a single
crystalline domain. In particular, the instability towards clumping
takes place only on large enough length scales.  In a polycrystalline
sample, if the size of the crystallites is too small, terms from the
elastic energy in (\ref{eom}) could dominate instead. In addition, it
is important that the sedimentation be steady, a requirement best met
by working in the fluidised-bed geometry in which the particles
constituting the crystal are on average at rest in the laboratory
frame of reference, and the fluid flows vertically upwards past
them. We would recommend starting with the suspension in the fully
sedimented state, and then switching on the upward flow.  Observations
in \cite{rutgersthesis} suggest that strongly charge-stabilised
crystalline suspensions appear stable whereas suspensions in a
fluid state display the Crowley instability in a visible manner. We
suspect that the instability is present even in the crystalline
suspensions, but is masked either by finite crystallite size or by the
logarithmically slow coarsening of domains.  We predict that careful
measurements of the time-evolution of the static structure factor,
using particle-imaging or ultrasmall-angle scattering techniques,
should reveal a weak large-scale modulation of the particle
concentration, with characteristic wavevector normal to the
sedimentation direction and decreasing logarithmically in time.

\section{Acknowledgement}
We are grateful to Deepak Dhar and Ramakrishna Ramaswamy for 
useful discussions.

\end{document}